\documentclass[%
 reprint,
superscriptaddress,
 amsmath,amssymb,
 aps,
physrev,
floatfix,
]{revtex4-2}

\usepackage{graphicx}
\usepackage{dcolumn}
\usepackage{bm}
\usepackage[
  colorlinks=true,
  linkcolor=blue,
  citecolor=blue,
  urlcolor=blue
]{hyperref}

\begin{document}


\title{Isometrization of Tensor Network States via Gauge Propagation}

\author{Zhiyu Jiang}
\email{jiang.zhiyu.qiqb@osaka-u.ac.jp}
\affiliation{Center for Quantum Information and Quantum Biology, The University of Osaka, Toyonaka, Osaka, 560-8531, Japan}


\author{Hiroshi Ueda}
\email{ueda.hiroshi.qiqb@osaka-u.ac.jp}
\affiliation{Center for Quantum Information and Quantum Biology, The University of Osaka, Toyonaka, Osaka, 560-8531, Japan}
\affiliation{Computational Materials Science Research Team, RIKEN Center for Computational Science (R-CCS), Kobe, Hyogo, 650-0047, Japan}





\begin{abstract}
We introduce a gauge-propagation approach for approximately converting generic tensor-network states into an isometric tensor-network form with a prescribed orthogonality center.
In one dimension, this propagation is exact because the non-isometric factor produced by a QR or singular-value decomposition is supported on a single virtual bond.
In higher-dimensional networks, however, a local step can have several outgoing directions, and the residual factor is generally not separable into independent single-bond contributions.
We address this local obstruction by approximating a local tensor, or a contracted local cluster, by structured terms consisting of an isometric factor multiplied by a tensor product of output-leg factors.
The isometric factor is retained at the current site or cluster, while the output-leg factors are absorbed into neighboring tensors along the propagation directions.
This construction applies to general local input-output partitions for which the input-side dimension is no smaller than the output-side dimension and provides a local truncation criterion for gauge propagation.
Benchmarks on random tensors show that the proposed decomposition is effective in the low-term regime and that the residual decreases for larger local clusters.
For the loop-gas tensor representation of the Kitaev spin liquid, two structured terms reduce the local residual to numerical precision, and the same cluster refinement further lowers the leading-term truncation error and reduces error accumulation during gauge propagation on a finite honeycomb network.
These results identify a propagation-compatible local decomposition as a useful building block for approximate isometrization and as a potential initializer or preconditioner for variational isoTNS algorithms.
\end{abstract}

\maketitle


\section{\label{sec:intro}Introduction}

Tensor-network representations provide a compact language for quantum many-body states and a practical foundation for numerical algorithms~\cite{orus2014practical,Orus2019TensorNetworks,Cirac2021Concepts}.
In one dimension, matrix product states (MPS), together with the density matrix renormalization group (DMRG)~\cite{PhysRevLett.69.2863,White1993DensityMatrix,RevModPhys.77.259,schollwock2011density}, provide controlled methods whose efficiency is closely tied to canonical forms and local gauge transformations~\cite{schollwock2011density,Fannes1992Finitely,Ostlund1995Thermodynamic,PerezGarcia2007MPS}.
This success is also connected to one-dimensional area-law structure~\cite{Hastings2007AreaLaw} and to efficient time-evolution and variational algorithms~\cite{Vidal2003Efficient,Vidal2004Efficient,Haegeman2011TDVP,Haegeman2016Unifying} within the MPS manifold.

Higher-dimensional tensor-network states, including tensor product states and projected entangled-pair states (PEPS), extend this variational framework to two-dimensional systems~\cite{verstraete2004renormalization,verstraete2008matrix,Jordan2008ClassicalSimulation}.
This extension is not straightforward: exact contraction and generic variational optimization of PEPS are computationally hard in general~\cite{PhysRevLett.98.140506,Markov2008Simulating,Corboz2016Variational,Vanderstraeten2016Gradient}.
This difficulty has motivated approximate contraction schemes, variational update strategies, gauge-fixing procedures, and entanglement-renormalization methods such as the multiscale entanglement renormalization ansatz (MERA)~\cite{Nishino1996CTMRG,Nishino1997CTM,Levin2007TRG,Gu2008TERG,Jiang2008Accurate,Orus2009Corner,Corboz2010Fermionic,Phien2015FastFullUpdate,Corboz2016Variational,Vanderstraeten2016Gradient,PhysRevLett.99.220405,PhysRevLett.101.110501,Evenbly2009Algorithms,Evenbly2015TNR}.

Isometric tensor-network states (isoTNS) provide a complementary route by imposing directional isometric constraints on higher-dimensional tensor networks.
These constraints retain part of the computational structure that makes one-dimensional canonical forms useful, including directional orthogonality and efficient contraction along selected paths~\cite{PhysRevLett.124.037201,lin2022efficient,Dai2025FermionicIsoTNS}.
A basic task is therefore isometrization, by which we mean the construction of an exact or approximate representation in which all tensors away from a chosen orthogonality center satisfy prescribed local isometric constraints.

The one-dimensional case explains both the appeal and the limitation of this idea.
For MPS, a canonical form can be constructed by sweeping QR or singular-value decompositions toward the orthogonality center.
At each step, the non-isometric factor produced by the decomposition acts on a single virtual bond and can be absorbed exactly into the neighboring tensor.
This single-bond structure is the reason that gauge propagation is exact in one dimension, apart from any intentional truncation.

In higher-dimensional networks, the same local step can have several outgoing directions.
Once an orthogonality center and a propagation ordering are fixed, a residual factor produced at a local tensor need not be a single-bond object.
For a strictly local propagation step, the residual would have to separate into factors that can be absorbed independently by neighboring tensors along the outgoing directions.
A generic tensor does not have this separability.
The smallest equal-dimensional obstruction is a 2-in-2-out local tensor, equivalently a two-qubit operator after grouping input and output legs.

Existing isoTNS algorithms address related difficulties through variational local updates.
For example, the Moses move redistributes entanglement while preserving the prescribed isometric structure~\cite{PhysRevLett.124.037201}.
Subsequent work has developed dynamical, infinite-strip, fermionic, and variational isoTNS algorithms~\cite{lin2022efficient,Wu_PRB_2023_isoTNS,Dai2025FermionicIsoTNS,Wu_PRXQ_2025_isoTNS}.
The computational complexity of isoTNS contractions has also been analyzed separately~\cite{Malz_PRXQ_2025_isoTNS}.
A diagonal isometric form with auxiliary tensors on the orthogonality hypersurface has recently provided another route and extends naturally to honeycomb and kagome geometries~\cite{Sappler2026DiagonalIsoTNS}.
These approaches are powerful, but they do not isolate the local algebraic obstruction that appears when one attempts to propagate gauge factors explicitly through a branching network.
Here we focus on this local obstruction and construct a decomposition tailored to it.

We introduce a propagation-compatible local decomposition in which a local tensor, or a contracted local cluster, is approximated by structured terms composed of an isometric factor and a tensor product of output-leg factors.
The isometric factor remains on the current tensor or effective cluster, while the output-leg factors are absorbed into neighboring tensors along the propagation directions.
Thus, the non-isometric content is reorganized into locally absorbable pieces rather than removed.
The construction applies to general local tensors or contracted clusters with a prescribed input-output orientation, provided that the input-side dimension is no smaller than the output-side dimension.
The minimal 2-in-2-out tensor is the smallest nontrivial case; enlarged 4-in-2-out and 6-in-2-out clusters provide a practical refinement by incorporating more of the short-range environment before truncation.

The leading coefficient of this decomposition measures the best Frobenius overlap with a normalized propagation-compatible term.
Keeping this leading term gives the elementary approximation used in the local gauge-propagation step, while additional terms can be extracted from the residual when higher local accuracy is required.
We derive a dimension-dependent lower bound on the optimal leading coefficient in terms of the trace norm.
For Frobenius-normalized complex Ginibre tensors, this bound further implies that the optimized one-term residual vanishes as the input dimension becomes large relative to the output dimension.

We test the method on random local tensors and on the loop-gas (LG) tensor-network representation of the Kitaev spin liquid (KSL)~\cite{kitaev2006anyons,Savary2017QuantumSpinLiquids,Hermanns2018Kitaev,Knolle2019FieldGuide,PhysRevLett.123.087203}.
For random tensors, the proposed decomposition gives smaller low-term residuals than operator-Schmidt and sorted Pauli reference truncations in the tested ensemble, while enlarging the local cluster improves the leading-term approximation.
Additional benchmarks on independent complex Ginibre tensors with different output-leg factorizations show that, over the tested range, the leading residual decreases as the input dimension grows relative to the output dimension.
For the LG tensor, two propagation-compatible terms reduce the local residual to numerical precision, while larger clusters further improve the leading-term approximation.
We further apply the decomposition to inward gauge propagation on a finite honeycomb LG tensor network and find that the same cluster refinement reduces accumulated state errors.

The method is intended as a local building block for approximate isometrization rather than as a replacement for variational isoTNS sweeps.
Its advantages are locality, explicit gauge propagation, and a transparent refinement parameter given by the number of retained terms or cluster size.
It may therefore be useful as a fast approximate isometrization procedure, as an initializer for subsequent isoTNS sweeps, or as a preconditioning step before more expensive variational refinement.

The remainder of the paper is organized as follows.
Section~\ref{sec:gauge_propagation} formulates gauge propagation, local isometric constraints, the orthogonality center, and the propagation ordering, and then identifies the obstruction beyond the one-dimensional exact limit.
Section~\ref{sec:local_decompositions} introduces the structured local decomposition, derives bounds for the leading-term approximation, and describes the optimization procedure used in the numerical calculations.
Section~\ref{sec:numerics} presents random-tensor benchmarks across cluster sizes and input-output dimensions, LG local decompositions, and finite honeycomb gauge propagation.
Section~\ref{sec:summary} summarizes the main conclusions and limitations.

\section{\label{sec:gauge_propagation}Gauge propagation and the isometrization problem}

In this section, we formulate the isometrization problem from the viewpoint of gauge propagation. 
The gauge freedom of tensor-network representations allows local degrees of freedom to be redistributed along internal bonds, while the isoTNS structure imposes isometric constraints with respect to chosen directions. 
This formulation requires a target orthogonality center and a corresponding propagation ordering. 
In one dimension, the resulting procedure is exact: gauge factors can be propagated by successive QR or singular-value decompositions. 
In higher dimensions, however, the remaining non-isometric part need not be confined to a single bond and may instead involve several neighboring directions.
These localized obstructions motivate the decomposition problem considered below.

\subsection{\label{subsec:gauge_constraints_ordering}Gauge freedom, isometric constraints, and propagation ordering}

Tensor-network representations contain a local gauge redundancy on their internal bonds.
On any such bond, an invertible matrix and its inverse may be inserted without changing the contracted many-body state.
For two neighboring tensors $T_i$ and $T_j$ contracted over a shared virtual index, this freedom can be written schematically as
\begin{equation}
T_i T_j = (T_i G)(G^{-1}T_j),
\label{eq:gauge_freedom}
\end{equation}
where $G$ acts on the shared virtual space.
The transformation changes the local tensor representatives assigned to the two sites while leaving the fully contracted network invariant.
Gauge propagation is the systematic use of this redundancy to move non-isometric degrees of freedom through the network.

The local constraints are formulated by regarding each tensor as a linear map.
For a tensor $T_v$ at site $v$, a chosen propagation direction induces a bipartition of its indices into incoming and outgoing sets, denoted by $I_v$ and $O_v$, respectively. 
For a tensor carrying a physical index, this index is included in the incoming set $I_v$ in our convention.
These sets define the input-side and output-side Hilbert spaces,
\begin{equation}
\mathcal H_{\rm in}(v)
=
\bigotimes_{\ell\in I_v}\mathcal H_\ell, \qquad
\mathcal H_{\rm out}(v)
=
\bigotimes_{\ell\in O_v}\mathcal H_\ell .
\label{eq:local_hilbert_spaces}
\end{equation}
After grouping the corresponding indices, the tensor is reshaped into a matrix representation of a linear map,
\begin{equation}
M_v:\mathcal H_{\rm out}(v)\to \mathcal H_{\rm in}(v).
\label{eq:local_tensor_map}
\end{equation}
With this convention, the local isometric constraint takes the form
\begin{equation}
M_v^\dagger M_v = I_{\mathcal H_{\rm out}(v)} .
\label{eq:local_isometry}
\end{equation}
Thus, the columns of $M_v$ are orthonormal; equivalently, $M_v$ preserves inner products on $\mathcal H_{\rm out}(v)$.
This condition is stronger than normalization of the tensor entries.
It constrains the tensor as a map between two specified Hilbert spaces and depends on the chosen input-output partition of the tensor indices. 

The isometrization problem can now be stated in these terms.
Given a tensor-network state represented by local tensors \(\{T_v\}\), one seeks an isometrized representation \(\{\widetilde T_v\}\) with a chosen orthogonality center \(c\) such that, for every \(v\neq c\), the matrix representation of \(\widetilde T_v\) associated with the prescribed input-output partition satisfies Eq.~\eqref{eq:local_isometry}.
The resulting state is required to be identical to the original state in the exact case, or close to it in an approximate construction,
\begin{equation}
|\Psi(\widetilde T)\rangle \simeq |\Psi(T)\rangle .
\label{eq:isometrization_target}
\end{equation}
The center tensor $\widetilde T_c$ is not constrained to be isometric and carries the remaining non-isometric degrees of freedom.

\begin{figure}[t]
\centering
\includegraphics[width=\columnwidth]{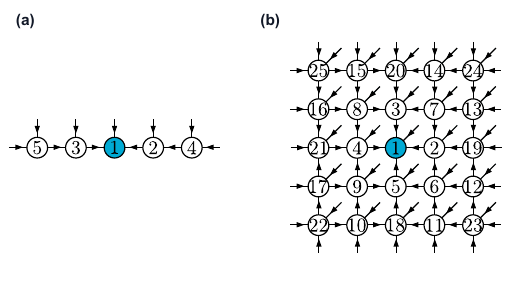}
\caption{
Schematic illustration of the gauge-propagation ordering.
The highlighted tensor labeled \(1\) is the orthogonality center, and arrows indicate gauge-propagation directions.
(a) One-dimensional chain with inward propagation from both sides.
(b) Two-dimensional lattice.
Away from the central row and column, a local step can have multiple outgoing propagation directions.
}
\label{fig:propagation_ordering}
\end{figure}

The same convention is used for all network structures considered in this work.
After choosing an orthogonality center $c$, we label it as $v_1$ and assign increasing labels to tensors farther from the center.
In the finite examples below, the labels are generated by increasing graph distance from the center, with a fixed geometric tie-breaking rule for sites at the same distance.
Gauge propagation follows the reverse order
\begin{equation}
v_N \to v_{N-1} \to \cdots \to v_2 \to v_1=c .
\label{eq:propagation_ordering}
\end{equation}
This convention is illustrated schematically in Fig.~\ref{fig:propagation_ordering}.
At the step associated with $v_k$, the remaining non-isometric factor is passed toward tensors with smaller labels.
This ordering determines the input-output partition of the tensor at $v_k$, and hence fixes which instance of the local isometric condition in Eq.~\eqref{eq:local_isometry} is imposed.

The propagation ordering is therefore part of the isometrization setup, not merely a convention for naming sites.
It fixes the local matrix representations on which the isoTNS constraints are imposed.
As shown in the following subsections, this distinction is immaterial in one dimension but becomes a source of localized obstructions in higher dimensions.

\subsection{\label{subsec:oned_svd}One-dimensional limit: exact SVD propagation}

The one-dimensional case provides the simplest realization of the propagation ordering defined above, as shown in Fig.~\ref{fig:propagation_ordering}(a).
Once an orthogonality center is chosen, non-isometric factors are propagated from sites farther from the center toward the center.
At each local step, the remaining non-isometric factor is passed to the neighboring tensor closer to the orthogonality center.
The propagation therefore has a unique local destination.

Consider a bulk MPS tensor $A_i^{s_i}$ with physical index $s_i$ and two virtual indices.
Following the convention of Sec.~\ref{subsec:gauge_constraints_ordering}, the physical index and the virtual index farther from the orthogonality center are grouped into a composite row index, while the virtual index closer to the center is grouped into a composite column index.
We denote the resulting matrix representation again by $A_i$.
A singular-value decomposition gives
\begin{equation}
A_i
=
U_i S_i V_i^\dagger ,
\label{eq:mps_svd}
\end{equation}
with
\begin{equation}
U_i^\dagger U_i=I .
\label{eq:mps_local_isometry}
\end{equation}
The isometric factor $U_i$ is assigned to site $i$, so that the local isometric constraint is satisfied at that site.
The remaining factor,
\begin{equation}
R_i=S_iV_i^\dagger,
\label{eq:mps_residual_factor}
\end{equation}
acts only on the virtual bond connecting site $i$ to the neighboring site closer to the orthogonality center.

The residual factor can therefore be absorbed exactly into that neighboring tensor $A_j$,
\begin{equation}
A_i A_{j}
=
U_i\left(R_i A_{j}\right),
\label{eq:mps_absorption}
\end{equation}
where $j$ denotes the neighboring site closer to the orthogonality center.
Repeating the same step propagates the non-isometric factor along the chain until it reaches the orthogonality center.

The exactness of this procedure relies on the absence of branching.
In one dimension, the residual factor produced at each step lives on a single virtual bond and has only one neighboring tensor into which it can be absorbed.
In the absence of truncation, gauge propagation by SVD is therefore exact in one dimension.
This single-bond structure is the feature that ceases to be automatic in higher-dimensional tensor networks.

\subsection{\label{subsec:higher_obstructions}Localized obstructions in higher dimensions}

In higher-dimensional networks, the single-bond structure identified above is no longer guaranteed.
As illustrated in Fig.~\ref{fig:propagation_ordering}(b), the boundary between the processed and unprocessed regions can involve several adjacent directions.
A local propagation step may then produce a residual object connected to more than one neighboring tensor.
Unlike the one-dimensional residual factor, such an object cannot, in general, be assigned to a single virtual bond.

For gauge propagation to remain local, this residual object must separate into factors associated with individual outgoing directions.
In that case, each factor can be absorbed independently into the corresponding neighboring tensor, reducing the step to a set of independent one-dimensional absorption steps.
The obstruction arises when such separation is impossible.
The non-isometric content is then distributed over several directions and cannot be localized onto individual bonds by a single product factorization.

In the two-direction case, the required separation can be written schematically as
\begin{equation}
R=X^{(1)}\otimes X^{(2)},
\label{eq:directional_product_factorization}
\end{equation}
where \(X^{(1)}\) and \(X^{(2)}\) are associated with the two outgoing directions.
A generic local tensor, however, does not admit such a product form.
For a general \(m\)-in-\(n\)-out partition with \(n>1\), the same question arises in the dimensionally allowed regime,
\begin{equation}
\dim\left(
\bigotimes_{a=1}^{m}\mathcal H_{I_a}
\right)
\ge
\dim\left(
\bigotimes_{b=1}^{n}\mathcal H_{O_b}
\right).
\label{eq:local_isometry_dimension_condition}
\end{equation}
If all index spaces in Eq.~\eqref{eq:local_isometry_dimension_condition} have the same dimension \(D>1\), this condition reduces to \(m\ge n\).
The smallest nontrivial equal-dimensional case with more than one outgoing direction is therefore \(m=n=2\), corresponding to a 2-in-2-out local tensor.

The obstruction is local and structural rather than merely numerical.
This is the point at which the one-dimensional SVD propagation mechanism ceases to apply directly.
The following section addresses this obstruction by introducing a local decomposition in which the remaining non-isometric factors are kept locally absorbable.

\section{\label{sec:local_decompositions}Localized decompositions for gauge propagation}

In this section, we develop the local decomposition used to approximate the obstruction identified above.
The main idea is to rewrite a local tensor as a sum of structured terms, each consisting of an isometric factor and local factors that can be absorbed into neighboring tensors.
This structure gives a local gauge-propagation step: the isometric factor remains on the current tensor or effective cluster, while the local factors are passed to adjacent tensors.
Within this class, truncating the decomposition to dominant terms gives a local approximation whose error is controlled by the residual.
We also summarize the optimization procedure used to construct these terms in the numerical calculations.

\subsection{\label{subsec:local_decomposition_structure}Local decomposition structure and normalization}

We first define the admissible local class.
All input-output partitions follow the convention of Sec.~\ref{sec:gauge_propagation}.
Here, a local tensor denotes either a single-site tensor or an effective tensor obtained by contracting a small local cluster.
After grouping its indices according to the propagation direction, we use the same symbol \(T\) for the local tensor and for its matrix representation.
The aim is to decompose \(T\) into terms whose non-isometric parts remain locally absorbable.

We write each structured local term as
\begin{equation}
\mathcal{F}_k = U_k X_k,
\qquad
X_k=\bigotimes_{b=1}^{n}X_k^{(b)} .
\label{eq:structured_term}
\end{equation}
Here \(U_k\) is the isometric part retained on the local tensor or effective cluster, while \(X_k^{(b)}\) acts only on the \(b\)-th outgoing leg group.
The factor \(U_k\) satisfies \(U_k^\dagger U_k=I\) and becomes unitary when the input and output dimensions are equal.
With the map convention used in Sec.~\ref{sec:gauge_propagation}, the product factor \(X_k\) acts on the outgoing side of the local map.
The factors \(X_k^{(b)}\) can then be absorbed into neighboring tensors along the corresponding outgoing directions.

The local decomposition of \(T\) is written as
\begin{equation}
T
\simeq
\sum_{k=1}^{K}\alpha_k \mathcal{F}_k
=
\sum_{k=1}^{K}
\alpha_k
U_k
\left(
\bigotimes_{b=1}^{n}X_k^{(b)}
\right).
\label{eq:local_decomposition_general}
\end{equation}
The symbol \(\simeq\) allows for truncation.
When all required terms are retained, the same expression represents an exact local decomposition.
In practical gauge propagation, however, one typically retains only a small number of dominant terms.

For the minimal equal-dimensional two-output case, Eq.~\eqref{eq:structured_term} reduces to the 2-in-2-out form
\begin{equation}
\mathcal{F}_k
=
U_k
\left(
A_k\otimes B_k
\right),
\label{eq:two_in_two_out_term}
\end{equation}
where \(A_k\) and \(B_k\) are the two local factors corresponding to \(X_k^{(1)}\) and \(X_k^{(2)}\), respectively.
Effective 4-in-2-out and 6-in-2-out tensors are treated as enlarged two-output local tensors, with \(U_k\) understood as a rectangular isometry.

We next fix the normalization used to define the coefficients \(\alpha_k\).
We use the Frobenius norm \(\|A\|_F=\sqrt{\langle A,A\rangle_F}\), where \(\langle A,B\rangle_F\) denotes the Frobenius inner product for tensors with the same index structure, equivalently \(\langle A,B\rangle_F=\operatorname{Tr}(A^\dagger B)\) after grouping indices into a matrix representation.
The local tensor is normalized as
\begin{equation}
\|T\|_F=1.
\label{eq:local_tensor_normalization}
\end{equation}
Each structured local term is also normalized as
\begin{equation}
\|\mathcal{F}_k\|_F=1.
\label{eq:term_normalization}
\end{equation}
Since \(U_k\) is isometric, this condition fixes the overall scale of the product factor \(X_k\).
Equivalently, the overall amplitude of each retained structured term is carried by the coefficient \(\alpha_k\).

The phase of \(\mathcal{F}_k\) is chosen so that \(\alpha_k\) is real and nonnegative.
With this convention, the scale of each retained contribution is carried by \(\alpha_k\), while \(\mathcal F_k\) specifies its normalized structured direction.
This normalization will be used in the next subsection to define the leading term and the error introduced by truncating the local decomposition.

\subsection{\label{subsec:leading_term_truncation}Approximation and leading-term truncation}

We now use the structured local decomposition as an approximation scheme.
With the normalization in Eqs.~\eqref{eq:local_tensor_normalization} and \eqref{eq:term_normalization}, the coefficients \(\alpha_k\) quantify the amplitudes of the normalized structured local terms.
When discussing an \(m\)-term truncation, we label the retained terms by decreasing coefficient,
\begin{equation}
\alpha_1\ge \alpha_2\ge \cdots \ge \alpha_K\ge 0 .
\label{eq:coefficient_ordering}
\end{equation}
With this convention, \(\mathcal{F}_1\) is the leading structured term.
The first term used in the local propagation step is obtained by maximizing the overlap with the original local tensor \(T\); subsequent terms, when used, are extracted from residual tensors.
This distinction is important because the leading term is the practical truncation used in the propagation calculations below.

The simplest approximation keeps only the leading structured term,
\begin{equation}
T^{(1)}
=
\alpha_1 \mathcal{F}_1 .
\label{eq:leading_term_approximation}
\end{equation}
In the gauge-propagation picture, this approximation gives a single local propagation channel.
The isometric factor in \(\mathcal{F}_1\) remains on the current tensor or effective cluster, while the local factors are absorbed into neighboring tensors along the outgoing directions.
Thus, the leading-term approximation gives a basic local step for the propagation procedure.

More generally, retaining the first \(m\) terms gives
\begin{equation}
T^{(m)}
=
\sum_{k=1}^{m}\alpha_k\mathcal{F}_k ,
\label{eq:m_term_truncation}
\end{equation}
with residual
\begin{equation}
R^{(m)}
=
T-T^{(m)} .
\label{eq:truncation_residual}
\end{equation}
If the decomposition in Eq.~\eqref{eq:local_decomposition_general} is exact before truncation, then any truncation after \(m\) retained normalized terms obeys
\begin{equation}
\|R^{(m)}\|_F
\le
\sum_{k=m+1}^{K}\alpha_k .
\label{eq:tail_bound}
\end{equation}
This bound does not require the structured terms \(\mathcal{F}_k\) to be mutually orthogonal.
It follows from the normalization \(\|\mathcal{F}_k\|_F=1\) and the triangle inequality.
The tail of an exact coefficient expansion therefore gives a conservative upper bound on the local truncation error.
For the greedy residual construction used in the numerical work, we report the residual norm directly.

The leading coefficient also has a variational interpretation.
For a normalized local tensor \(T\), let \(\mathcal C\) denote the class of normalized structured local terms defined by Eqs.~\eqref{eq:structured_term} and \eqref{eq:term_normalization}.
The best one-term approximation within this class is characterized by
\begin{equation}
\alpha_1^{\rm opt}
=
\max_{\mathcal F\in \mathcal C}
\left|
\langle \mathcal F,T\rangle_F
\right| .
\label{eq:leading_coefficient_opt}
\end{equation}
The phase of the maximizing term is chosen so that the overlap is real and nonnegative.
For the optimized leading term \(\mathcal{F}_1\), the corresponding projection error is
\begin{equation}
\|T-\alpha_1^{\rm opt}\mathcal{F}_1\|_F^2
=
1-\left(\alpha_1^{\rm opt}\right)^2 .
\label{eq:single_term_projection_error}
\end{equation}
Thus, a larger leading coefficient gives a smaller one-term local error.

The optimized leading coefficient can be bounded from below by restricting the variational search in Eq.~\eqref{eq:leading_coefficient_opt} to structured terms with identity output-side factors.

Let
\begin{equation}
D_{\rm out}
=
\dim \mathcal H_{\rm out}
=
\prod_{b=1}^{n} d_b ,
\label{eq:output_dimension}
\end{equation}
where \(d_b\) is the dimension of the \(b\)-th outgoing leg group.
For any isometry \(U\) with the same input-output partition as \(T\), consider the structured term
\begin{equation}
\mathcal{F}_U
=
U
\left(
\bigotimes_{b=1}^{n}
\frac{I_b}{\sqrt{d_b}}
\right).
\label{eq:identity_factor_embedding_general}
\end{equation}
Each identity factor is Frobenius-normalized, and their tensor product is \(I_{\rm out}/\sqrt{D_{\rm out}}\).
Therefore,
\begin{equation}
\mathcal{F}_U
=
\frac{1}{\sqrt{D_{\rm out}}}U .
\label{eq:identity_factor_scaled_isometry}
\end{equation}
Since \(U^\dagger U=I_{\rm out}\), one has
\(\|U\|_F^2=\operatorname{Tr}(U^\dagger U)=D_{\rm out}\), and hence
\(\|\mathcal F_U\|_F=1\).
Thus, \(\mathcal F_U\) belongs to the admissible class \(\mathcal C\).

Restricting the maximization in Eq.~\eqref{eq:leading_coefficient_opt} to this subset gives
\begin{equation}
\alpha_1^{\rm opt}
\ge
\max_{U^\dagger U=I_{\rm out}}
\left|
\langle \mathcal F_U,T\rangle_F
\right|.
\label{eq:identity_factor_bound_general_start}
\end{equation}
Substituting Eq.~\eqref{eq:identity_factor_scaled_isometry}, we obtain
\begin{equation}
\alpha_1^{\rm opt}
\ge
\frac{1}{\sqrt{D_{\rm out}}}
\max_{U^\dagger U=I_{\rm out}}
\left|
\operatorname{Tr}
\left(
U^\dagger T
\right)
\right|.
\label{eq:identity_factor_bound_step}
\end{equation}
The maximum over isometries is equal to the trace norm of \(T\),
\begin{equation}
\max_{U^\dagger U=I_{\rm out}}
\left|
\operatorname{Tr}
\left(
U^\dagger T
\right)
\right|
=
\|T\|_* ,
\label{eq:trace_norm_variational}
\end{equation}
where \(\|T\|_*\) is the sum of the singular values of \(T\).
It follows that
\begin{equation}
\alpha_1^{\rm opt}
\ge
\frac{\|T\|_*}{\sqrt{D_{\rm out}}}.
\label{eq:trace_norm_bound_general}
\end{equation}
Because \(T\) is Frobenius-normalized and \(\|T\|_*\geq\|T\|_F=1\), this immediately gives
\begin{equation}
\alpha_1^{\rm opt}
\ge
\frac{1}{\sqrt{D_{\rm out}}}.
\label{eq:leading_coefficient_general_bound}
\end{equation}

For the two-qubit 2-in-2-out case, the two outgoing legs each have dimension \(2\), so that \(D_{\rm out}=4\).
Equation~\eqref{eq:leading_coefficient_general_bound} then gives
\begin{equation}
\alpha_1^{\rm opt}
\ge
\frac{1}{2}.
\label{eq:leading_coefficient_bound}
\end{equation}
This provides a worst-case reference for the leading-term approximation in the minimal 2-in-2-out setting.

The stronger trace-norm bound in Eq.~\eqref{eq:trace_norm_bound_general} also yields an asymptotic result for tall random matrices.
Let \(T\in\mathbb C^{D_{\rm in}\times D_{\rm out}}\), where \(D_{\rm in}\) and \(D_{\rm out}\) are the input-side and output-side dimensions, respectively, and suppose that \(T\) is obtained by Frobenius-normalizing a complex Ginibre matrix \(G\),
\begin{equation}
T
=
\frac{G}{\|G\|_F},
\qquad
T^\dagger T
=
\frac{G^\dagger G}
{\operatorname{Tr}(G^\dagger G)}.
\label{eq:normalized_ginibre_tensor}
\end{equation}
For a tall Gaussian matrix, the columns become nearly orthogonal and have comparable norms as the input-side dimension increases.
In the limit \(D_{\rm out}/D_{\rm in}\to0\), standard results for the extreme singular values of rectangular Gaussian matrices imply that, with probability approaching unity~\cite{davidson2001local},
\begin{equation}
T^\dagger T
\longrightarrow
\frac{I_{D_{\rm out}}}{D_{\rm out}} .
\label{eq:tall_random_gram_limit}
\end{equation}
The same behavior can also be understood from the small-aspect-ratio limit of the Marchenko--Pastur distribution~\cite{marchenko1967distribution}.

The singular values of \(T\) therefore approach \(1/\sqrt{D_{\rm out}}\).
Consequently,
\begin{equation}
\frac{\|T\|_*}{\sqrt{D_{\rm out}}}
\longrightarrow
\frac{1}{\sqrt{D_{\rm out}}}
\sum_{a=1}^{D_{\rm out}}
\frac{1}{\sqrt{D_{\rm out}}}
=
1 .
\label{eq:tall_random_trace_norm_limit}
\end{equation}
Combining this result with Eq.~\eqref{eq:trace_norm_bound_general} and the normalization bound \(\alpha_1^{\rm opt}\leq1\), we obtain
\begin{equation}
\alpha_1^{\rm opt}
\longrightarrow
1 .
\label{eq:leading_coefficient_tall_limit}
\end{equation}
Thus, a single normalized propagation-compatible term asymptotically captures the entire Frobenius weight of the local tensor.
Equivalently, using Eq.~\eqref{eq:single_term_projection_error}, the corresponding optimized one-term residual satisfies
\begin{equation}
\left\|
T-\alpha_1^{\rm opt}\mathcal F_1
\right\|_F
=
\sqrt{
1-\left(\alpha_1^{\rm opt}\right)^2
}
\longrightarrow
0 .
\label{eq:leading_term_error_tall_limit}
\end{equation}
Therefore, for random matrices whose input-side dimension is much larger than their output-side dimension, the leading-term approximation becomes asymptotically exact.

This result provides a random-matrix perspective on the cluster-size dependence examined below.
Contracting neighboring tensors into an effective 4-in-2-out or 6-in-2-out cluster increases the input-side dimension while leaving the output-side dimension unchanged, thereby reducing the aspect ratio \(D_{\rm out}/D_{\rm in}\).
This dimensional change moves the effective tensor toward the same dimensional regime analyzed above.

Contracted cluster tensors generally contain correlations and additional structure and need not be Ginibre distributed, so the random-matrix result does not establish a monotonic improvement for arbitrary clusters.
It nevertheless identifies a plausible dimension-driven mechanism for improved leading-term accuracy and motivates the localized cluster hierarchy examined numerically below.
Increasing the cluster size then trades a larger local optimization problem for a potentially smaller leading-term truncation error.

\subsection{\label{subsec:optimization_algorithm}Term extraction used in the numerical implementation}

For completeness, we describe the term-extraction procedure used in the numerical calculations.
The central step is the extraction of a leading term from a given tensor.
The same step is then applied recursively to residual tensors to obtain a multi-term decomposition.

\subsubsection{\label{subsubsec:leading_term_extraction}Leading-term extraction}

The leading-term extraction is applied to a current tensor \(R\), which is either the original local tensor \(T\) or a residual generated during the iterative decomposition.
The task is to find the normalized structured term with the largest Frobenius overlap with \(R\),
\begin{equation}
\alpha(R)
=
\max_{\mathcal F\in\mathcal C}
\left|
\langle \mathcal F,R\rangle_F
\right| .
\label{eq:algorithm_leading_problem}
\end{equation}
The optimized term is denoted by \(\mathcal F\), and the corresponding coefficient is \(\alpha(R)\).

For the update rules, we parameterize a trial element \(\mathcal F\in\mathcal C\) as in Eq.~\eqref{eq:structured_term},
\begin{equation}
\mathcal F
=
U X,
\qquad
X=
\bigotimes_{b=1}^{n}X^{(b)} .
\label{eq:algorithm_trial_term}
\end{equation}
The optimization is performed by alternating between the isometric factor \(U\) and the product factor \(X\).
The overall phase of \(\mathcal F\) is fixed at the end so that \(\langle \mathcal F,R\rangle_F\) is real and nonnegative.

For fixed local factors \(X^{(b)}\), the update of \(U\) is an isometric Procrustes problem~\cite{schonemann1966procrustes,higham1986polar}.
Using \(X=\bigotimes_b X^{(b)}\), define
\begin{equation}
M
=
R X^\dagger .
\label{eq:algorithm_procrustes_matrix}
\end{equation}
Then the \(U\)-update is
\begin{equation}
\max_{U^\dagger U=I}
\operatorname{Re}
\left\langle
U,
M
\right\rangle_F .
\label{eq:algorithm_procrustes_problem}
\end{equation}
If the thin singular-value decomposition of \(M\) is
\begin{equation}
M
=
P\Sigma Q^\dagger ,
\label{eq:algorithm_procrustes_svd}
\end{equation}
the update is
\begin{equation}
U
\leftarrow
P Q^\dagger .
\label{eq:algorithm_procrustes_update}
\end{equation}
This update gives an isometric polar factor of \(M\) and satisfies \(U^\dagger U=I\).

For fixed \(U\), the update of the local factors is a product-factor approximation.
Define
\begin{equation}
Y
=
U^\dagger R .
\label{eq:algorithm_product_target}
\end{equation}
The product-factor problem is
\begin{equation}
\max_{\{X^{(b)}\}}
\operatorname{Re}
\left\langle
Y,
\bigotimes_{b=1}^{n}X^{(b)}
\right\rangle_F ,
\label{eq:algorithm_product_problem}
\end{equation}
with the normalization convention of Eq.~\eqref{eq:term_normalization}.
We solve this problem by updating one factor at a time.
For a fixed \(b\), all factors \(X^{(c)}\) with \(c\neq b\) are held fixed, and we define the effective local target \(Y_{\rm eff}^{(b)}\) by
\begin{equation}
\left\langle
Y,
\bigotimes_{c=1}^{n}X^{(c)}
\right\rangle_F
=
\left\langle
Y_{\rm eff}^{(b)},
X^{(b)}
\right\rangle_F .
\label{eq:algorithm_effective_target}
\end{equation}
The optimal update of the \(b\)-th factor is then
\begin{equation}
X^{(b)}
\leftarrow
\frac{
Y_{\rm eff}^{(b)}
}{
\|Y_{\rm eff}^{(b)}\|_F
}.
\label{eq:algorithm_factor_update}
\end{equation}
The separate normalization of the local factors fixes their scale freedom.
Because \(U\) is isometric and \(\|X\|_F=\prod_{b=1}^{n}\|X^{(b)}\|_F=1\), the resulting structured term automatically satisfies \(\|\mathcal F\|_F=1\).

For the two-output tensors used in the following, this product-factor update is a nearest-Kronecker-product approximation in the Frobenius norm, which can be written as a best rank-one approximation after realignment~\cite{VanLoan1993,DeLathauwer2000Multilinear,VanLoan2000Kronecker,kolda2009tensor,Oseledets2011TensorTrain}.
In this case \(X=A\otimes B\), we fix the remaining scale freedom by normalizing the two local factors separately.
The product-factor problem in Eq.~\eqref{eq:algorithm_product_problem} then becomes
\begin{equation}
\max_{A,B}
\operatorname{Re}
\left\langle
Y,
A\otimes B
\right\rangle_F ,
\qquad
\|A\|_F=\|B\|_F=1 .
\label{eq:two_output_product_problem}
\end{equation}
We define the realignment map \(\mathcal R\) by grouping the indices associated with the two outgoing factors,
\begin{equation}
\left[\mathcal R(Y)\right]_{(i,j),(k,l)}
=
Y_{(i,k),(j,l)} .
\label{eq:two_output_realignment}
\end{equation}
With this realignment, the overlap with \(A\otimes B\) is equivalent to the overlap of \(\mathcal R(Y)\) with a rank-one matrix built from \(\operatorname{vec}(A)\) and \(\operatorname{vec}(B)\).
If
\begin{equation}
\mathcal R(Y)
=
\sum_r s_r x_r y_r^\dagger
\label{eq:two_output_realignment_svd}
\end{equation}
is a singular-value decomposition, the optimal two-output product update is obtained from the leading singular vectors,
\begin{equation}
A
\leftarrow
\operatorname{unvec}(x_1),
\qquad
B
\leftarrow
\operatorname{unvec}(y_1^{*}),
\label{eq:two_output_product_update}
\end{equation}
followed by Frobenius normalization.
Here \(\operatorname{unvec}\) denotes the inverse of the vectorization map used in the realignment.
This is the product-factor update used for the two-output local tensors in the numerical calculations below, including the 2-in-2-out, 4-in-2-out, and 6-in-2-out cases.

Each block update optimizes the objective over one subset of variables while keeping the others fixed, so the objective is nondecreasing under exact updates.
The alternating procedure is not guaranteed to find the global optimum.
In the calculations below, we therefore use several random initializations and retain the solution with the largest overlap.

\subsubsection{\label{subsubsec:multi_term_decomposition}Iterative multi-term decomposition}

The same extraction step can be applied recursively to construct a finite-term approximation.
The resulting approximation to a normalized local tensor \(T\) has the form
\begin{equation}
T
\simeq
\sum_{k=1}^{K}
\alpha_k \mathcal F_k ,
\label{eq:algorithm_multiterm_target}
\end{equation}
where each \(\mathcal F_k\) is a normalized structured term and the coefficients are real and nonnegative.

We initialize the residual as
\begin{equation}
R^{(0)}
=
T .
\label{eq:algorithm_initial_residual}
\end{equation}
At step \(k\), the leading-term extraction is applied to the current residual \(R^{(k-1)}\).
This gives a normalized structured term \(\mathcal F_k\) and a coefficient
\begin{equation}
\alpha_k
=
\left\langle
\mathcal F_k,
R^{(k-1)}
\right\rangle_F
\ge
0 .
\label{eq:algorithm_greedy_coefficient}
\end{equation}
The residual is then updated by subtracting the extracted contribution,
\begin{equation}
R^{(k)}
=
R^{(k-1)}
-
\alpha_k \mathcal F_k .
\label{eq:algorithm_residual_update}
\end{equation}
This iteration may be stopped either after a prescribed number of terms or when the residual norm falls below a chosen local tolerance \(\tau_{\rm tol}\),
\begin{equation}
\left\|R^{(k)}\right\|_F \leq \tau_{\rm tol}.
\label{eq:algorithm_stopping_condition}
\end{equation}

After \(K\) steps, the decomposition takes the form
\begin{equation}
T
=
\sum_{k=1}^{K}
\alpha_k \mathcal F_k
+
R^{(K)} .
\label{eq:algorithm_greedy_decomposition}
\end{equation}
Equivalently, the \(K\)-term truncation is
\begin{equation}
T^{(K)}
=
\sum_{k=1}^{K}
\alpha_k \mathcal F_k ,
\label{eq:algorithm_K_term_truncation}
\end{equation}
with residual error \(\|R^{(K)}\|_F\).

This procedure is greedy because each term is chosen to maximize the overlap with the current residual.
It gives a practical local construction of the structured decomposition, but it does not generally produce the globally optimal \(K\)-term approximation.
The first extracted term is the leading term discussed in Sec.~\ref{subsec:leading_term_truncation}, while subsequent terms provide controlled multi-term refinements when higher local accuracy is required.

\section{\label{sec:numerics}Numerical analysis}

In this section, we test the propagation-compatible decomposition on random local tensors and on LG tensor-network states with physical structure.
Random local tensors provide a structure-free baseline because no algebraic or physical structure is available to assist the decomposition.
This benchmark isolates the local approximation power of the method, with particular emphasis on the leading-term truncation used in local gauge propagation.
We then consider the LG tensor-network representation of the KSL, for which the physical structure of the local tensor is expected to make the decomposition more efficient.
In both settings, we compare localized cluster choices, denoted 2-in-2-out, 4-in-2-out, and 6-in-2-out, to assess the accuracy gained by enlarging the local propagation region.
The same hierarchy is then tested in inward gauge propagation on a finite LG tensor network as a proof-of-concept benchmark.
The results show that the proposed decomposition gives an effective local truncation in the tested settings, captures the LG local tensor with only a few terms, and reduces the accumulated propagation error as the local cluster is enlarged.

\subsection{\label{subsec:local_benchmarks}Local decomposition benchmarks}

\begin{figure}[tb]
\centering
\includegraphics[width=\columnwidth]{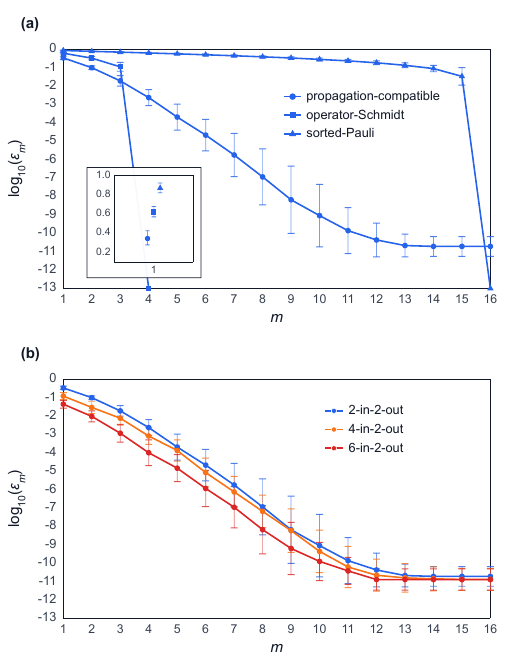}
\caption{
Local decomposition benchmarks for random tensors with leg dimension \(\chi=2\).
(a) Comparison between the propagation-compatible decomposition, the operator-Schmidt decomposition, and the sorted Pauli expansion for random 2-in-2-out tensors.
(b) Cluster-size comparison using 2-in-2-out, 4-in-2-out, and 6-in-2-out local clusters.
Here \(m\) is the number of retained terms and \(\varepsilon_m\) is the Frobenius residual after retaining \(m\) terms.
Curves show \(\langle \log_{10}\varepsilon_m\rangle\) over 20 samples, and error bars indicate one sample standard deviation of \(\log_{10}\varepsilon_m\).
}
\label{fig:random_local_benchmarks}
\end{figure}

We begin with random local tensors in the minimal 2-in-2-out setting, taking each leg dimension to be \(\chi=2\).
After grouping the two incoming and two outgoing legs, each sample is represented as a $4\times4$ complex matrix.
The entries are drawn from a complex Ginibre ensemble, and the resulting matrix is normalized with respect to the Frobenius norm.
We denote the normalized tensor by $\widehat T$.
The data in Fig.~\ref{fig:random_local_benchmarks} use 20 independent samples, eight random initializations for each leading-term extraction, and a maximum of 120 alternating updates per initialization.
The plotted curves show sample means of \(\log_{10}\varepsilon_m\), while the error bars indicate one sample standard deviation.

For this benchmark, we use the truncation convention introduced in Sec.~\ref{subsec:leading_term_truncation}.
Since each sampled tensor is Frobenius-normalized before decomposition, the plotted error is the Frobenius norm of the residual,
\begin{equation}
\label{eq:random_residual_error}
\varepsilon_m
=
\left\|
\widehat{T}
-
\sum_{\mu=1}^{m}
\alpha_{\mu}\mathcal F_{\mu}
\right\|_F .
\end{equation}
The case $m=1$ gives the leading-term truncation used in the most local implementation of gauge propagation.
The decay of $\varepsilon_m$ with increasing $m$ indicates how efficiently the residual can be reduced by retaining additional terms.

In the 2-in-2-out case, the tensor can be viewed as a two-qubit operator after grouping the two incoming and two outgoing legs~\cite{NielsenChuang2010Quantum}.
This allows a comparison between the propagation-compatible decomposition and two standard reference decompositions.
For the chosen grouping of the two local input-output pairs, the operator-Schmidt decomposition gives a basis-independent bipartite product expansion~\cite{Tyson_2003},
\begin{equation}
\label{eq:random_operator_schmidt}
\widehat{T}
=
\sum_{\nu=1}^{r}
s_{\nu}
\left(
A_{\nu}\otimes B_{\nu}
\right),
\qquad
\langle A_{\mu},A_{\nu}\rangle_F
=
\langle B_{\mu},B_{\nu}\rangle_F
=
\delta_{\mu\nu}.
\end{equation}
The corresponding $m$-term truncation retains the $m$ largest singular values $s_{\nu}$.
This provides a natural bipartite product reference, although the bipartition is not determined by the gauge-propagation geometry.

We also use a sorted Pauli expansion in a fixed orthonormal operator basis.
Writing $\widetilde{\sigma}_0=I/\sqrt{2}$ and taking $\widetilde{\sigma}_{1,2,3}$ to be the normalized Pauli matrices, we expand
\begin{equation}
\label{eq:random_sorted_pauli}
\widehat{T}
=
\sum_{a,b=0}^{3}
c_{ab}
\left(
\widetilde{\sigma}_a\otimes\widetilde{\sigma}_b
\right),
\qquad
c_{ab}
=
\left\langle
\widetilde{\sigma}_a\otimes\widetilde{\sigma}_b,
\widehat T
\right\rangle_F .
\end{equation}
The sorted Pauli truncation keeps the $m$ largest coefficients $|c_{ab}|$.
This provides a complete operator-basis reference, although the fixed Pauli basis is not selected by the gauge-propagation geometry.

For an arbitrary two-qubit operator, the reference decompositions become exact once sufficiently many terms are retained: the operator-Schmidt decomposition is exact by $m=4$ for this bipartition, while the sorted Pauli expansion is exact by $m=16$.
In local gauge propagation, however, the full expansion is not the practical regime of interest, because retaining more terms increases the cost of the subsequent tensor-network manipulation.
The relevant comparison is therefore the low-term truncation regime.
As shown on the logarithmic scale in Fig.~\ref{fig:random_local_benchmarks}(a), the proposed decomposition gives smaller residual errors than the two references at small $m$ for the tested random ensemble.
This advantage is already visible at $m=1$, which is the most relevant leading-term approximation for local propagation because only a single propagation-compatible term is retained.

The leading-term result suggests a complementary way to improve the local approximation without increasing the number of retained terms.
Instead of increasing $m$, one can enlarge the local tensor before performing the decomposition.
In addition to the minimal 2-in-2-out tensor, we consider effective 4-in-2-out and 6-in-2-out local tensors, generated by contracting two or three random 2-in-2-out tensors along an output-leg chain.
The two uncontracted output legs define the outgoing directions of the enlarged local object.
At fixed $m$, a larger cluster incorporates more short-range structure into the local decomposition problem before truncation, at the cost of a larger local calculation.

Figure~\ref{fig:random_local_benchmarks}(b) shows the resulting cluster-size hierarchy for random tensors.
The residual error decreases as the local cluster is enlarged, with the reduction already visible in the leading-term truncation at $m=1$.
Together with Fig.~\ref{fig:random_local_benchmarks}(a), this shows that the proposed decomposition is effective in the low-term regime and can be systematically refined by increasing the local cluster size.

\begin{figure}[t]
\centering
\includegraphics[width=\columnwidth]{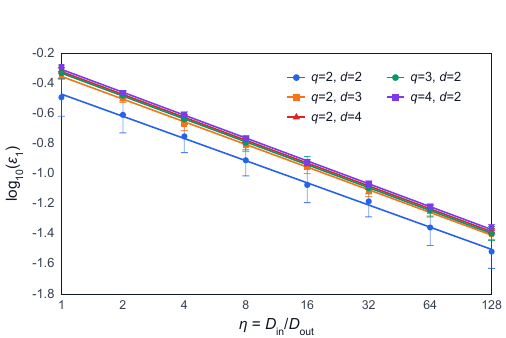}
\caption{
Leading-term benchmark for normalized complex Ginibre tensors with different output-leg factorizations.
Here \(q\) is the number of outgoing legs, \(d\) is their common dimension, and
\(\eta=D_{\rm in}/D_{\rm out}\).
The data cover \(\eta=1,2,4,\ldots,128\).
Symbols denote sample means of \(\log_{10}\varepsilon_1\) over 20 random tensors, error bars show one sample standard deviation in the same logarithmic variable, and solid lines are least-squares fits to Eq.~\eqref{eq:empirical_aspect_scaling}.
}
\label{fig:random_aspect_ratio}
\end{figure}

The cluster comparison above preserves the two-output propagation structure and improves the approximation by enlarging the local object.
More generally, however, the structured class in Eq.~\eqref{eq:local_decomposition_general} is not restricted to two outgoing leg groups or to qubit dimensions.
It applies to any input-output partition satisfying Eq.~\eqref{eq:local_isometry_dimension_condition}, with one locally absorbable factor associated with each outgoing leg group.
The minimal 2-in-2-out tensor is therefore only the smallest nontrivial instance of the general construction.

To quantify this broader dimensional dependence, we consider normalized complex Ginibre tensors with \(q\) outgoing leg groups of common dimension \(d\), total output dimension \(D_{\rm out}=d^q\), and fused input dimension \(D_{\rm in}=\eta D_{\rm out}\).
Retaining only the leading propagation-compatible term, we obtain the aspect-ratio dependence shown in Fig.~\ref{fig:random_aspect_ratio}.
For each fixed output structure \((q,d)\), we first fit the data to the generic power law \(\varepsilon_1\simeq C\eta^{-\gamma}\), allowing both \(C\) and \(\gamma\) to vary independently.
The fitted exponents lie in the range \(0.4895\leq\gamma\leq0.5052\), cluster closely around \(1/2\), and exhibit no discernible systematic dependence on the tested output structures.
By contrast, the fitted prefactors vary appreciably across these structures.
We therefore denote the prefactor by \(C_{\rm fit}(q,d)\) and summarize the observed scaling as
\begin{equation}
\varepsilon_1
\simeq
C_{\rm fit}(q,d)\eta^{-1/2}.
\label{eq:empirical_aspect_scaling}
\end{equation}
A theoretical reference for the \(\eta^{-1/2}\) scale and a further empirical analysis of the output-structure dependence are given in Appendix~\ref{app:aspect_ratio_scaling}.

These random-tensor benchmarks demonstrate the applicability of the propagation-compatible construction beyond the minimal two-qubit local setting.
We next consider the LG tensor, where the local object carries additional algebraic structure from the underlying physical model.

\subsection{\label{subsec:loopgas_local}LG local tensor decompositions}

We next apply the propagation-compatible decomposition introduced in Sec.~\ref{sec:local_decompositions} to a tensor with physical constraints.
As a representative structured example, we consider the LG tensor-network representation of the KSL~\cite{kitaev2006anyons,PhysRevLett.123.087203}.
This tensor is defined directly in the spin basis and encodes the vortex-free constraint together with the underlying $\mathbb{Z}_2$ gauge structure.
It therefore provides a structured counterpart to the random-tensor results of Sec.~\ref{subsec:local_benchmarks} and allows us to examine whether the propagation-compatible decomposition can exploit model-specific structure in a physically motivated tensor.

The LG tensor network is defined on the honeycomb lattice, where each site carries a physical spin-$1/2$ index and three virtual indices.
Denoting the virtual indices by $i,j,k\in\{0,1\}$, the local tensor is a spinor in the physical spin space,
\begin{equation}
\left|T_{ijk}\right\rangle
=
\sum_s
T^s_{ijk}
\left|s\right\rangle .
\label{eq:loopgas_tensor_spinor}
\end{equation}
In the calculations below, we use the zeroth-order LG tensor,
\begin{equation}
\left|T^{(0)}_{ijk}\right\rangle
=
\hat Q_{ijk}
\left|(111)\right\rangle ,
\label{eq:loopgas_tensor_definition}
\end{equation}
where $\hat Q_{ijk}$ has matrix elements
\begin{equation}
Q^{ss'}_{ijk}
=
\tau_{ijk}
\left[
(\sigma^x)^{1-i}
(\sigma^y)^{1-j}
(\sigma^z)^{1-k}
\right]_{ss'} .
\label{eq:loopgas_Q_definition}
\end{equation}
The only nonzero values of $\tau_{ijk}$ are
\begin{equation}
\tau_{000}=-\mathrm{i},
\qquad
\tau_{011}=\tau_{101}=\tau_{110}=1 .
\label{eq:loopgas_tau_definition}
\end{equation}
The reference spinor $\left|(111)\right\rangle$ is polarized along the $(1,1,1)$ direction,
\begin{equation}
\left\langle (111)\right|
\boldsymbol{\sigma}
\left|(111)\right\rangle
=
\frac{1}{\sqrt{3}}(1,1,1) .
\label{eq:loopgas_reference_spinor}
\end{equation}
In the $\sigma^z$ basis, we write
\begin{equation}
\left|(111)\right\rangle
=
a_\uparrow\left|\uparrow\right\rangle
+
a_\downarrow\left|\downarrow\right\rangle ,
\label{eq:loopgas_spinor_cd}
\end{equation}
where
\begin{equation}
a_\uparrow=
\sqrt{\frac{1+1/\sqrt{3}}{2}},
\qquad
a_\downarrow=
e^{\mathrm{i}\pi/4}
\sqrt{\frac{1-1/\sqrt{3}}{2}} .
\label{eq:loopgas_cd_definition}
\end{equation}

To connect this tensor with the two-output local decomposition used in Sec.~\ref{sec:local_decompositions}, we choose a 2-in-2-out partition consistent with the convention of Sec.~\ref{subsec:gauge_constraints_ordering}.
The physical index is included in the incoming set.
We therefore group the physical index $s$ with one virtual index, chosen here to be $i$, and assign the remaining virtual indices $j$ and $k$ to the two outgoing legs.
With this input-output grouping, the local tensor is represented as
\begin{equation}
T^{(0)}_{(s,i),(j,k)}
=
\begin{pmatrix}
a_\uparrow & 0 & 0 & a_\downarrow \\
a_\downarrow & 0 & 0 & a_\uparrow \\
0 & -\mathrm{i}a_\downarrow & a_\uparrow & 0 \\
0 & \mathrm{i}a_\uparrow & -a_\downarrow & 0
\end{pmatrix}.
\label{eq:loopgas_2in2out_matrix}
\end{equation}
As in the random-tensor analysis, this local matrix is Frobenius-normalized before applying the decomposition.
We denote the resulting normalized 2-in-2-out LG tensor by $\widehat T_{\rm LG}^{(2\to2)}$.

\begin{figure}[tb]
\centering
\includegraphics[width=\columnwidth]{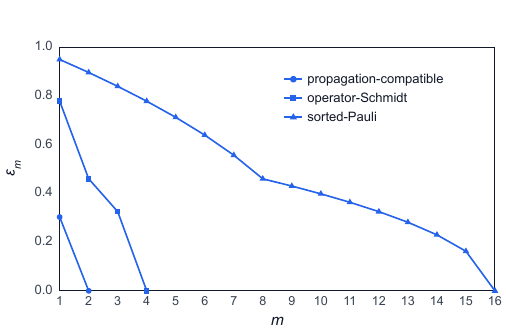}
\caption{
Local decomposition benchmark for the minimal 2-in-2-out LG tensor $\widehat T_{\rm LG}^{(2\to2)}$.
The propagation-compatible decomposition is compared with the operator-Schmidt decomposition and the sorted Pauli expansion.
Here $m$ denotes the number of retained terms, and $\varepsilon_m$ denotes the Frobenius residual error after retaining $m$ terms.
}
\label{fig:loopgas_2in2out_benchmark}
\end{figure}

For $\widehat T_{\rm LG}^{(2\to2)}$, the propagation-compatible decomposition exhibits an even stronger low-term structure than the random-tensor benchmark.
As shown in Fig.~\ref{fig:loopgas_2in2out_benchmark}, retaining only two propagation-compatible terms reduces the residual to numerical precision.
Thus, the local LG tensor is captured to numerical precision by a two-term truncation within the proposed decomposition class.
A clear advantage is already visible in the leading-term truncation, where the $m=1$ residual is smaller than those obtained from the operator-Schmidt decomposition and the sorted Pauli expansion.
This rapid convergence reflects the algebraic structure of the LG local tensor and indicates that the admissible output-leg class is well matched to this physical tensor.

Although the two-term result establishes the accuracy of the local decomposition, the leading-term regime remains the most relevant one for a minimal gauge-propagation step.
We next assess whether the leading local truncation can be further improved by enlarging the tensor before decomposition.
To this end, we construct effective two-output tensors by exactly contracting one or two neighboring LG tensors into the minimal 2-in-2-out object while keeping the two propagation directions open.
These contractions give the Frobenius-normalized tensors $\widehat T_{\rm LG}^{(4\to2)}$ and $\widehat T_{\rm LG}^{(6\to2)}$, respectively.
In both enlarged clusters, the two propagation directions are retained as outgoing directions, and the corresponding uncontracted boundary virtual legs define the outgoing legs of the effective tensor.
All remaining open physical and virtual indices are grouped into the input side.
Thus, the three tensors have the same two-output structure but contain different amounts of local environment before the approximation is made, realizing the localized-cluster hierarchy discussed in Sec.~\ref{subsec:leading_term_truncation}.

\begin{table}[!t]
\caption{
Cluster-size comparison for LG local tensor decompositions using 2-in-2-out, 4-in-2-out, and 6-in-2-out local clusters.
The entries give the Frobenius residual error $\varepsilon_m$ after retaining $m$ propagation-compatible terms for each cluster size.
}
\label{tab:loopgas_cluster_errors}
\begin{ruledtabular}
\begin{tabular}{cccc}
$m$ & 2-in-2-out & 4-in-2-out & 6-in-2-out \\
\colrule
\noalign{\vskip 2pt}
1 & $3.03\times 10^{-1}$ & $1.69\times 10^{-1}$ & $9.67\times 10^{-2}$ \\
2 & $1.95\times 10^{-15}$ & $1.73\times 10^{-15}$ & $1.47\times 10^{-15}$ \\
\end{tabular}
\end{ruledtabular}
\end{table}

The cluster-size comparison is summarized in Table~\ref{tab:loopgas_cluster_errors}.
The main effect of enlarging the local cluster appears in the leading-term truncation.
At $m=1$, the residual error decreases from $3.03\times 10^{-1}$ for the 2-in-2-out tensor to $1.69\times 10^{-1}$ and $9.67\times 10^{-2}$ for the 4-in-2-out and 6-in-2-out clusters, respectively.
Thus, cluster enlargement reduces the dominant local truncation error.
The $m=2$ row shows that two terms are sufficient to reach numerical precision for the enlarged clusters as well.

In these enlarged-cluster calculations, the isometric factor is treated as an effective cluster tensor. Restoring a fixed single-site tensor-network geometry would require an additional factorization or variational projection step, which is beyond the scope of the present local benchmark.

These results show that the propagation-compatible decomposition efficiently captures the local structure of the LG tensor, and that enlarging the local cluster reduces the dominant truncation error.
For tensor-network isometrization, however, the local approximation is applied repeatedly as the gauge is propagated through the network.
The relevant question is therefore how the resulting local errors accumulate along a propagation path.
We address this question in the next subsection using a finite honeycomb LG tensor network with an orthogonality center.

\subsection{\label{subsec:loopgas_disk}Gauge propagation on a finite honeycomb LG tensor network}

We finally apply the local decomposition hierarchy to gauge propagation in a finite tensor network.
The purpose is to examine how the local errors introduced by leading-term truncation of the propagation-compatible decomposition accumulate when the decomposition is applied repeatedly.
We consider a finite honeycomb LG tensor network and choose an orthogonality center.
The tensors are labeled according to the propagation ordering introduced in Sec.~\ref{subsec:gauge_constraints_ordering}, with the gauge propagated inward toward the center.

\begin{figure}[tb]
\centering
\includegraphics[width=\columnwidth]{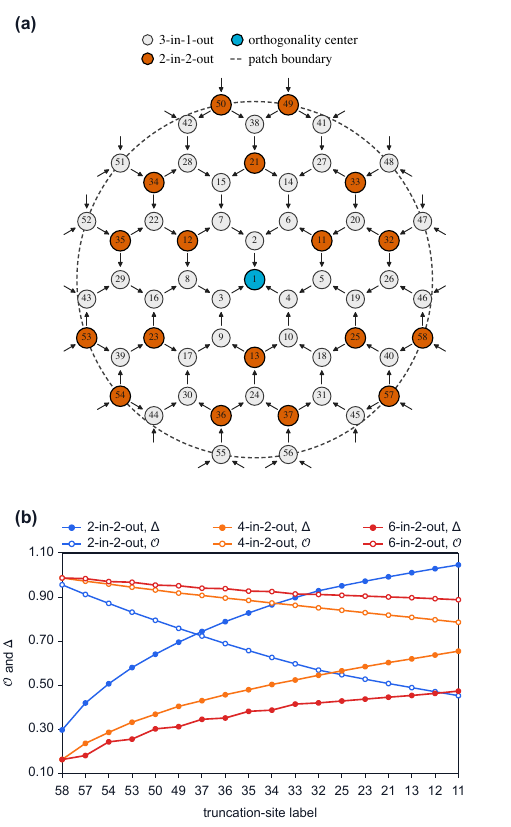}
\caption{
Gauge propagation on a finite honeycomb LG tensor network containing 58 physical sites and 24 open auxiliary boundary indices, with physical indices omitted.
The omitted physical indices are included in the incoming index groups under the convention of Sec.~\ref{subsec:gauge_constraints_ordering}.
(a) Propagation geometry toward the orthogonality center.
Sites are labeled by the inward propagation order; 2-in-2-out sites mark the truncation events, while 3-in-1-out sites are treated by exact single-output SVD steps.
(b) Accumulated normalized state error $\Delta_\ell$ and normalized overlap $\mathcal O_\ell$ during inward propagation.
The horizontal-axis labels in (b) correspond to the 2-in-2-out truncation sites in (a).
For the curve labeled 6-in-2-out, full 6-in-2-out clusters are used when available, while constrained sites are treated by the largest available two-output cluster with the same outgoing propagation directions.
}
\label{fig:loopgas_network_propagation}
\end{figure}

The propagation geometry is shown in Fig.~\ref{fig:loopgas_network_propagation}(a).
In the present honeycomb ordering, the 3-in-1-out sites have only one outgoing virtual direction.
They are therefore treated by an exact single-output SVD step, in which the non-isometric factor is associated with a single outgoing bond and is absorbed into the neighboring tensor closer to the orthogonality center.
The nontrivial approximation occurs at the 2-in-2-out sites.
At such a site, the residual generally has support on two outgoing directions and cannot be assigned to a single bond.
We first treat these sites using the minimal 2-in-2-out propagation-compatible decomposition and retain only its leading term.

To quantify the accumulated error, we compare the propagated approximate state with the original finite-network state after each two-output truncation event.
We write $\Psi_{\rm exact}$ for the state represented by the original finite LG tensor network.
For a two-output truncation event labeled by $\ell$, the resulting approximate state is denoted by $\Psi_\ell$.
The labels $\ell$ correspond to the site-order labels of the 2-in-2-out sites shown in Fig.~\ref{fig:loopgas_network_propagation}(a) and used on the horizontal axis of Fig.~\ref{fig:loopgas_network_propagation}(b).
The finite patch is treated as a tensor with 58 physical spin-1/2 indices and 24 open auxiliary boundary indices, and the Frobenius inner products below are taken over all open indices.

We define the normalized overlap
\begin{equation}
\mathcal O_\ell
=
\frac{
\left|
\left\langle
\Psi_{\rm exact},
\Psi_\ell
\right\rangle_F
\right|
}{
\left\|\Psi_{\rm exact}\right\|_F
\left\|\Psi_\ell\right\|_F
}.
\label{eq:loopgas_network_overlap}
\end{equation}
The accumulated propagation error is then measured by the phase-insensitive chordal distance,
\begin{equation}
\Delta_\ell
=
\sqrt{ 2 \left( 1 - \mathcal O_\ell \right) }.
\label{eq:loopgas_network_state_error}
\end{equation}
This distance is closely related to standard fidelity-based distances for pure states~\cite{Jozsa1994}.

The result for the minimal 2-in-2-out scheme is shown in Fig.~\ref{fig:loopgas_network_propagation}(b).
As inward propagation proceeds and the leading-term truncation is applied at successive 2-in-2-out sites, the normalized state error $\Delta_\ell$ increases and the normalized overlap $\mathcal O_\ell$ decreases.
This trend shows the accumulation of errors generated by the minimal local decomposition during repeated gauge propagation.

To reduce the accumulated error seen in the minimal 2-in-2-out propagation, we next apply the enlarged local clusters introduced in Sec.~\ref{subsec:loopgas_local} along the same propagation sequence.
At each 2-in-2-out truncation site, the local object is enlarged before decomposition by contracting one or two neighboring tensors in the current network whenever available, giving the 4-in-2-out and 6-in-2-out propagation schemes, respectively.
The resulting isometric factor is retained as an effective cluster tensor during the subsequent propagation.
When the full enlarged cluster cannot be formed near the patch boundary, the largest available two-output cluster with the same outgoing propagation directions is used.
The finite network, orthogonality center, and truncation-site sequence are kept fixed, so that the comparison isolates the effect of the local cluster size.

As shown in Fig.~\ref{fig:loopgas_network_propagation}(b), enlarging the local cluster reduces the accumulated propagation error in this finite-network benchmark.
At the same truncation-site labels, the 4-in-2-out scheme gives a smaller $\Delta_\ell$ and a larger $\mathcal O_\ell$ than the minimal 2-in-2-out scheme.
The curve labeled 6-in-2-out gives the lowest accumulated error overall, with the corresponding largest normalized overlap.

The 6-in-2-out curve also exhibits an alternating pattern in both $\Delta_\ell$ and $\mathcal O_\ell$.
This pattern reflects the finite-network constraint on forming enlarged clusters.
A 4-in-2-out cluster can be formed at each 2-in-2-out truncation site by contracting one neighboring 3-in-1-out tensor.
By contrast, a full 6-in-2-out cluster requires two neighboring 3-in-1-out tensors.
Full 6-in-2-out clusters therefore cannot be assigned to every 2-in-2-out truncation site in the sequence because adjacent truncation sites compete for the same nearby 3-in-1-out tensors.
The curve labeled 6-in-2-out represents a largest-available enlarged-cluster scheme, in which full 6-in-2-out clusters are used where they can be formed and the remaining constrained sites are treated by 4-in-2-out clusters.
Under the chosen assignment, the first truncation site is treated by the 4-in-2-out fallback; this accounts for the coincidence of the initial 6-in-2-out and 4-in-2-out data points.
The alternating changes in $\Delta_\ell$ and $\mathcal O_\ell$ are therefore a direct consequence of this mixed 6-in-2-out/4-in-2-out assignment.

These results show how the residuals left by leading-term truncation accumulate when the propagation-compatible decomposition is applied repeatedly during finite-network isometrization.
The local improvement observed in Sec.~\ref{subsec:loopgas_local} is therefore reflected in the propagated state, not only in isolated tensor decompositions.
Together with the random-tensor and local LG analyses above, this finite-network test supports the propagation-compatible decomposition as a practical local building block for approximate gauge isometrization, with enlarged clusters providing a useful route for reducing the accumulated error.

\section{\label{sec:summary}Summary and outlook}

We have formulated approximate isometrization as a local gauge-propagation problem.
In one-dimensional tensor networks, the non-isometric factor produced by a local QR or SVD lives on a single bond and can be propagated exactly toward the orthogonality center.
In higher-dimensional networks, a local step may have several outgoing directions, and the residual factor is generally not a product of single-direction contributions.
This multi-output residual is the local obstruction addressed in this work.

Our proposed decomposition rewrites a local tensor, or a small contracted cluster, as structured terms consisting of an isometric factor and a tensor product of output-leg factors.
The isometric factor is kept at the current site or effective cluster, while the output-leg factors remain absorbable by neighboring tensors along the propagation directions.
The leading coefficient gives a local diagnostic for the best one-term propagation step, and in the minimal two-qubit 2-in-2-out case the optimized leading coefficient obeys a dimension-controlled lower bound of \(1/2\) under the Frobenius normalization used here.
Larger 4-in-2-out and 6-in-2-out clusters preserve the same two-output propagation structure while incorporating more of the local environment into the isometric part. 

The numerical tests support this local picture.
For random 2-in-2-out tensors, the propagation-compatible decomposition gives smaller low-term residuals than the operator-Schmidt and sorted Pauli reference truncations in the tested ensemble.
For independent Ginibre tensors with more general output-leg factorizations, the leading residual is consistent with the same \(\eta^{-1/2}\) aspect-ratio scaling over the tested range, while the output structure sets a finite-dimensional baseline prefactor.
For the LG tensor representation of the KSL, two structured terms reduce the isolated local residual to numerical precision.
In finite honeycomb gauge propagation, the accumulated state error decreases when the local object is enlarged from the minimal 2-in-2-out tensor to 4-in-2-out clusters and then to the largest-available scheme labeled 6-in-2-out.
The local cluster hierarchy therefore provides a practical refinement parameter for reducing the accumulated error generated by repeated leading-term truncations.

The method is complementary to variational isoTNS algorithms.
Sweep-based approaches, including local updates such as the Moses move~\cite{PhysRevLett.124.037201}, remain the natural route for high-accuracy variational optimization.
The gauge-propagation construction developed here is more local and explicit: once the orthogonality center and propagation ordering are fixed, each nontrivial step reduces to a local decomposition with a directly measurable residual.
It may therefore be useful as a fast approximate isometrization procedure, an initializer for later isoTNS sweeps, or a preconditioning step before more expensive variational refinement.

The 2-in-2-out case also has a direct circuit interpretation.
There the isometric factor is a two-qubit unitary, which can be related to standard two-qubit gate decompositions~\cite{Kraus_PRA_2001_dec,vatan2004optimal,Vidal2004Universal,Shende2004Minimal}, while the output-leg factors act as local tensor-network filters.
For enlarged 4-in-2-out and 6-in-2-out clusters, the isometric factor becomes a rectangular cluster isometry rather than a two-qubit unitary.
Such an isometry can be embedded into a unitary acting on the cluster degrees of freedom together with ancillary qubits, and explicit circuit representations may be obtained by general isometry-synthesis constructions. Approximate circuit encodings of the resulting cluster isometries may also be explored using quantum-circuit-encoding approaches such as automatic quantum circuit encoding~\cite{PhysRevA.93.032318,PhysRevResearch.6.043008}.
This viewpoint exposes a tradeoff between local approximation quality and circuit-level implementability: enlarged clusters improve tensor-network accuracy, but require additional ancillas and circuit-synthesis effort.
Circuit representations of the resulting isometries may also be useful as reference-state circuits in hybrid algorithms that combine tensor-network contractions with quantum-circuit overlaps, such as recent variational quantum SVD schemes using auxiliary reference states~\cite{Miyakoshi2026VQSVD}.
Assessing the optimal state-preparation cost and performance of this direction is left for future work.

Several limitations remain.
The alternating extraction algorithm is local and does not guarantee a globally optimal structured term.
The accuracy of leading-term propagation depends on the tensor structure, the propagation ordering, and the availability of enlarged clusters near finite-network boundaries.
The Frobenius residual used here provides a convenient local error measure, but sharper bounds are needed to relate local truncation errors to the global state fidelity.
A further practical issue is that enlarged-cluster decompositions produce effective cluster isometries; if one requires a fixed original tensor-network geometry, an additional factorization or variational projection step is needed.
Future work should address adaptive cluster selection, symmetry-resolved decompositions, circuit-aware cost functions, explicit restoration of fixed network geometry after cluster enlargement, and the use of the present construction inside full isoTNS optimization workflows.

\begin{acknowledgments}
This work was supported by JST CREST (No.~JPMJCR24I1), MEXT Q-LEAP (No.~JPMXS0120319794), JST COI-NEXT (No.~JPMJPF2014), and JSPS KAKENHI (No.~JP26K00664). 
It was also partially supported by the COE Research Grant in Computational Science from Hyogo Prefecture and Kobe City through Foundation for Computational Science.
\end{acknowledgments}

\appendix

\section{Aspect-ratio scaling and square-matrix baseline}
\label{app:aspect_ratio_scaling}

This Appendix provides a random-matrix reference for the leading propagation-compatible truncation beyond the minimal two-qubit setting.
We consider normalized complex Ginibre tensors with \(q\) outgoing leg groups, each of dimension \(d\).
The total outgoing dimension is
\begin{equation}
D_{\rm out}=d^q,
\end{equation}
while all incoming legs are fused into a space of dimension
\begin{equation}
D_{\rm in}=\eta D_{\rm out},
\qquad
\eta\geq1.
\label{eq:app_aspect_ratio}
\end{equation}
The leading structured term has one locally absorbable factor for each outgoing leg group, as in Eq.~\eqref{eq:local_decomposition_general}.

For a normalized tensor \(\widehat T\), the leading-term residual is
\begin{equation}
\varepsilon_1
=
\left\|
\widehat T-\alpha_1\mathcal F_1
\right\|_F
=
\sqrt{1-\alpha_1^2}.
\label{eq:app_leading_residual}
\end{equation}
The second equality follows from the variational optimization of \(\alpha_1\) when \(\mathcal F_1\) is normalized.
The aspect-ratio results in the main text are therefore equivalently statements about the approach of the dominant coefficient \(\alpha_1\) to unity.

\subsection{Random-matrix reference scale}

A useful admissible reference is obtained by choosing the normalized identity product factor on the outgoing space.
In the matrix convention of Sec.~\ref{sec:local_decompositions}, write the right polar decomposition as
\begin{equation}
\widehat T
=
U_{\rm pol}P,
\qquad
U_{\rm pol}^{\dagger}U_{\rm pol}
=
I_{D_{\rm out}},
\qquad
P
=
\left(
\widehat T^\dagger\widehat T
\right)^{1/2}.
\label{eq:app_right_polar}
\end{equation}
For a complex Ginibre matrix with \(D_{\rm in}\geq D_{\rm out}\), \(U_{\rm pol}\) is an isometry with probability one.

The normalized identity product factor is
\begin{equation}
P_{\rm id}
=
\frac{I_{D_{\rm out}}}{\sqrt{D_{\rm out}}}
=
\bigotimes_{b=1}^{q}
\frac{I_d}{\sqrt d}.
\label{eq:app_identity_factor}
\end{equation}
The factor \(P_{\rm id}\) belongs to the admissible output-leg product class.
Combining it with the polar isometry defines the structured term
\begin{equation}
\mathcal F_{\rm id}
=
U_{\rm pol}P_{\rm id}
=
\frac{1}{\sqrt{D_{\rm out}}}
U_{\rm pol}.
\label{eq:app_identity_structured_term}
\end{equation}
Because \(U_{\rm pol}\) is isometric and \(\|P_{\rm id}\|_F=1\), this term satisfies
\begin{equation}
\|\mathcal F_{\rm id}\|_F
=
1.
\end{equation}
Thus, \(\mathcal F_{\rm id}\) is an admissible, although not generally optimal, structured term.

Let \(G\in\mathbb C^{D_{\rm in}\times D_{\rm out}}\) be a complex Ginibre matrix and set
\begin{equation}
\widehat T
=
\frac{G}{\|G\|_F}.
\end{equation}
With
\begin{equation}
W
=
\frac{G^\dagger G}{D_{\rm in}},
\qquad
c
=
\frac{D_{\rm out}}{D_{\rm in}}
=
\eta^{-1},
\label{eq:app_wishart_definition}
\end{equation}
the overlap of \(\mathcal F_{\rm id}\) with \(\widehat T\) is exactly
\begin{equation}
\alpha_{\rm id}
\equiv
\left\langle
\mathcal F_{\rm id},
\widehat T
\right\rangle_F
=
\frac{\operatorname{Tr}\sqrt W}
{\sqrt{D_{\rm out}\operatorname{Tr}W}}.
\label{eq:app_identity_overlap_exact}
\end{equation}
This overlap is real and nonnegative because \(P\) is positive semidefinite.

Consider the joint large-dimension limit
\begin{equation}
D_{\rm in},D_{\rm out}
\to
\infty,
\qquad
\frac{D_{\rm out}}{D_{\rm in}}
\to
c
\in
(0,1].
\label{eq:app_joint_dimension_limit}
\end{equation}
The Marchenko--Pastur law implies that, in this joint limit,
the following quantities converge in probability:
\begin{equation}
\frac{\operatorname{Tr}W}{D_{\rm out}}
\longrightarrow
1,
\qquad
\frac{1}{D_{\rm out}}
\operatorname{Tr}\sqrt W
\longrightarrow
\int dx\,
\rho_c(x)\sqrt{x},
\end{equation}
where \(\rho_c\) is the Marchenko--Pastur density~\cite{marchenko1967distribution}.
Consequently,
\begin{equation}
\alpha_{\rm id}
\longrightarrow
\alpha_{\rm id}^{(\infty)}(c)
\equiv
\int dx\,
\rho_c(x)\sqrt{x}.
\label{eq:app_identity_overlap}
\end{equation}

Expanding the Marchenko--Pastur integral in the small-aspect-ratio regime \(c\ll1\) gives
\begin{equation}
\alpha_{\rm id}^{(\infty)}(c)
=
1-\frac{c}{8}+O(c^2).
\label{eq:app_identity_overlap_small_c}
\end{equation}
For each finite-dimensional realization, define the residual associated with the identity-product structured term by
\begin{equation}
\varepsilon_{\rm id}
\equiv
\left\|
\widehat T
-
\alpha_{\rm id}\mathcal F_{\rm id}
\right\|_F
=
\sqrt{1-\alpha_{\rm id}^2}.
\label{eq:app_identity_residual}
\end{equation}
Its large-dimension limit therefore satisfies
\begin{align}
\varepsilon_{\rm id}^{(\infty)}(c)
&\equiv
\sqrt{
1-
\left[
\alpha_{\rm id}^{(\infty)}(c)
\right]^2
}
\nonumber\\
&=
\frac{1}{2}c^{1/2}
+
O\!\left(c^{3/2}\right)
\nonumber\\
&=
\frac{1}{2}\eta^{-1/2}
+
O\!\left(\eta^{-3/2}\right).
\label{eq:app_identity_scaling}
\end{align}

Writing the globally optimized leading residual as
\begin{equation}
\varepsilon_1^{\rm opt}
=
\sqrt{
1-
\left(
\alpha_1^{\rm opt}
\right)^2
},
\end{equation}
the admissibility of \(\mathcal F_{\rm id}\) implies
\begin{equation}
\varepsilon_1^{\rm opt}
\leq
\varepsilon_{\rm id}.
\label{eq:app_optimized_identity_bound}
\end{equation}
Equation~\eqref{eq:app_identity_scaling} therefore provides a natural large-dimension reference scale for the leading residual.
It does not determine the prefactor of the optimized propagation-compatible result.

The numerical data in Fig.~\ref{fig:random_aspect_ratio} are consistent with this reference scale.
Independent fits to the generic power law \(\varepsilon_1\simeq C\eta^{-\gamma}\) give \(0.4895\leq\gamma\leq0.5052\) for the five sampled output structures.
Thus, over the tested range, the dominant variation with input redundancy is consistent with an \(\eta^{-1/2}\) scaling, while the output-leg structure primarily affects the finite-dimensional prefactor.

\subsection{Square-matrix baseline and empirical fit}

To isolate the dependence on the outgoing-side product structure, we define the square-matrix baseline as the geometric mean of the leading residual at \(\eta=1\).
\begin{equation}
\begin{aligned}
\ell_r(q,d)
&=
\log_{10}
\varepsilon_1^{(r)}(q,d,\eta=1),
\qquad
r=1,\ldots,N_s,
\\
C_0(q,d)
&=
10^{
\frac{1}{N_s}
\sum_{r=1}^{N_s}
\ell_r(q,d)
},
\qquad
N_s=20.
\end{aligned}
\label{eq:app_c0_definition}
\end{equation}
We sample the configurations \((q,d)=(2,2\text{--}8)\), \((3,2\text{--}4)\), \((4,2\text{--}3)\), \((5,2)\), and \((6,2)\), yielding 14 baseline values in total.
The sampled range becomes narrower at large \(q\) or \(d\) because the local optimization cost grows rapidly with \(D_{\rm out}=d^q\).

\begin{figure}[t]
\centering
\includegraphics[width=\columnwidth]{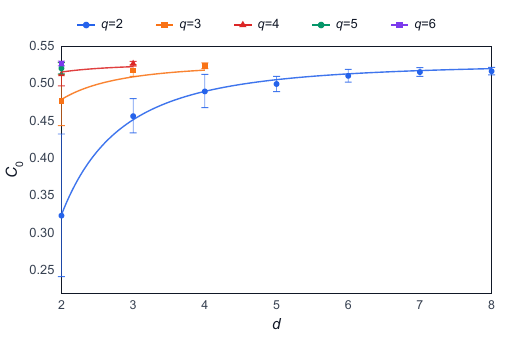}
\caption{
Square-matrix baseline \(C_0(q,d)\) for independent complex Ginibre tensors.
Symbols show the geometric mean of the leading residual at \(\eta=1\).
Error bars indicate the multiplicative one-standard-deviation interval
\(C_0(q,d)10^{\pm\sigma_\ell}\), where \(\sigma_\ell\) is the sample standard deviation of
\(\ell_r=\log_{10}\varepsilon_1^{(r)}\) over 20 samples.
Solid curves are obtained from Eq.~\eqref{eq:app_c0_fit} using the globally fitted parameters and are shown only over the sampled range of \(d\) for each \(q\).
The fitted parameters are \(C_\infty=0.5277(12)\), \(A=1.10(22)\), \(\lambda=1.44(13)\), and \(p=2.44(18)\), where parentheses denote bootstrap standard errors in the last quoted digits.
The \(q=5\) and \(q=6\) data sets each contain only the \(d=2\) point.
}
\label{fig:c0_scaling}
\end{figure}

We fit these 14 baseline values jointly using the empirical form
\begin{equation}
C_0^{\rm fit}(q,d)
=
C_\infty
-
A
e^{-\lambda(q-2)}
d^{-p}.
\label{eq:app_c0_fit}
\end{equation}
The parameters \(C_\infty\), \(A\), \(\lambda\), and \(p\) are determined simultaneously from this global fit.
The measured baselines and the resulting fit are shown in Fig.~\ref{fig:c0_scaling}.

We next compare the fitted asymptote with an independently derived random-matrix reference.
In the joint large-dimension square-matrix limit, \(D_{\rm in}=D_{\rm out}\) and hence \(c=1\).
Equation~\eqref{eq:app_identity_overlap} then yields
\begin{equation}
\alpha_{\rm id}^{(\infty)}(1)
=
\int_0^4
dx\,
\rho_1(x)\sqrt{x}
=
\frac{8}{3\pi}.
\end{equation}
The corresponding limiting residual of the identity-product reference is
\begin{equation}
C_\infty^{\rm id}
=
\sqrt{
1-
\left(
\frac{8}{3\pi}
\right)^2
}
=
0.52867.
\label{eq:app_cinfinity}
\end{equation}
This independently derived reference value agrees with the fitted \(C_\infty\) within one bootstrap standard error.
The agreement is consistent with the identity-product reference providing the large-dimension limiting scale.
It does not, however, establish that the optimized baseline \(C_0\) converges exactly to \(C_\infty^{\rm id}\), because the bootstrap uncertainty does not account for the choice of empirical fitting form or for extrapolation beyond the sampled parameter range.

Within the chosen empirical ansatz, the positive fitted values of \(\lambda\) and \(p\) imply that the finite-dimensional correction decreases as either the number of outgoing leg groups or their individual dimension increases.
The available data do not determine these two dependences independently, particularly at large \(q\), where only a limited range of \(d\) is sampled.
The large-\(q\) data therefore constrain the shared global form in Eq.~\eqref{eq:app_c0_fit} rather than a separate \(d\)-dependence for each \(q\).

Motivated by the common \(\eta^{-1/2}\) scaling, we use the composite estimate
\begin{equation}
\varepsilon_1(q,d,\eta)
\simeq
C_0^{\rm fit}(q,d)\eta^{-1/2}.
\label{eq:app_composite_estimate}
\end{equation}
Here \(C_0(q,d)\) is the directly measured \(\eta=1\) baseline, whereas \(C_0^{\rm fit}(q,d)\) is its empirical representation in Eq.~\eqref{eq:app_c0_fit}.
The finite-\(\eta\) intercept \(C_{\rm fit}(q,d)\) in Eq.~\eqref{eq:empirical_aspect_scaling} is obtained from a separate regression and need not coincide exactly with \(C_0(q,d)\).
Equation~\eqref{eq:app_composite_estimate} is therefore an empirical approximation rather than an exact identity.

Independent holdout tests on Ginibre tensors, not used in the fit, are performed at \((q,d)=(2,9)\), \((3,5)\), and \((7,2)\), with each configuration evaluated at \(\eta=1,4,16\).
The ratios of the observed geometric-mean residuals to the predictions of Eq.~\eqref{eq:app_composite_estimate} range from \(0.947\) to \(1.008\), with a root-mean-square logarithmic deviation of \(0.0169\) decades.
These tests support modest extrapolation of Eq.~\eqref{eq:app_composite_estimate} in \(q\) or \(d\) within the independent-Ginibre ensemble.

Equation~\eqref{eq:app_composite_estimate} is not intended as a universal law for structured or correlated tensors.
In particular, contracted random-tensor chains and the LG tensor contain correlations absent from the independent-Ginibre ensemble, and their leading residuals must be evaluated directly.

\bibliography{references}

\end{document}